\documentclass[pra,amssymb,amsmath,superscriptaddress,twocolumn,nofootinbib]{revtex4-1}

\usepackage{graphicx,color}
\usepackage[version=3]{mhchem}

\newcommand{\hl}[1]{\textcolor[RGB]{255,0,0}{#1}}
\renewcommand{\hl}[1]{#1} % remove highlighting by uncommenting this line

\graphicspath{{pix/}}

\begin{document}

\title[]{Model-independent measurements of the sodium magneto-optical trap's excited-state population}
\author{J.M. Kwolek}
\email{jonathan.kwolek@uconn.edu}
\affiliation{Department of Physics, University of Connecticut, Storrs, Connecticut 06268}

\author{D.S. Goodman}
\affiliation{Department of Sciences, Wentworth Institute of Technology, Boston, Massachusetts 02115}
\affiliation{Department of Physics, University of Connecticut, Storrs, Connecticut 06268}

\author{S.A. Entner}
\affiliation{Department of Sciences, Wentworth Institute of Technology, Boston, Massachusetts 02115}

\author{J.E. Wells}
\affiliation{W. M. Keck Science Department of Claremont McKenna, Pitzer, and Scripps Colleges, Claremont, California 91711}
\affiliation{Department of Physics, University of Connecticut, Storrs, Connecticut 06268}

\author{F.A. Narducci}
\affiliation{Department of Applied Physics, Naval Postgraduate School, Monterey, California 93943}

\author{W.W. Smith}
%\email{Winthrop.Smith@uconn.edu}
\affiliation{Department of Physics, University of Connecticut, Storrs, Connecticut 06268}
\date{\today}

\begin{abstract}
We present model-independent measurements of the excited-state population of atoms in a sodium (Na) magneto-optical trap (MOT) using a hybrid ion-neutral trap composed of a MOT and a linear Paul trap (LPT). We photoionize excited Na atoms trapped in the MOT and use two independent methods to measure the resulting ions: directly by trapping them in our LPT, and indirectly by monitoring changes in MOT fluorescence. By measuring the ionization rate via these two independent methods, we have enough information to directly determine the population of MOT atoms in the excited-state. The resulting measurement reveals that there is a range of trapping-laser intensities where the excited-state population of atoms in our MOT follows the standard two-level model intensity-dependence. However, an experimentally determined effective saturation intensity must be used instead of the theoretically predicted value from the two-level model. We measured the effective saturation intensity to be \hl{$I_\mathrm{se}=22.9\pm5.1\:\textrm{mW}/\textrm{cm}^2$ for the type-I Na MOT and $I_\mathrm{se}=49\pm11\;\textrm{mW}/\textrm{cm}^2$} for the type-II Na MOT, approximately 1.7 and 3.6 times the theoretical estimate, respectively. Lastly, at large trapping-laser intensities, our experiment reveals a clear departure from the two-level model at a critical intensity that we believe is due to a state-mixing effect, whose critical intensity can be determined by a simple power broadening model. 
\end{abstract}

\maketitle

\section{Introduction}

Magneto-optical traps (MOTs) are the workhorse of many modern cold atom experiments. A conventional MOT consists of six circularly-polarized beams of near-resonant light, oriented along the three Cartesian axes intersecting in a central trapping region. The light is detuned slightly from the atomic resonance of the atom (or sometimes molecule \cite{Demille:2010}), creating an optical molasses \cite{Lett:1989} and cooling the atom down by many orders of magnitude. The light force (due to momentum transfer from repeated absorption of near-resonant photons) has a spatial dependence given by a specially oriented magnetic-field gradient, confining the cold atoms to the center of the trapping region \cite{Raab:1987}. 

Accurate knowledge of the steady-state fraction of MOT atoms in the excited-state, $f_e$, is a critical and fundamental characterization of a MOT. For example, $f_e$ is traditionally used in determining the total number of atoms within a MOT \cite{Raab:1987}. In fact, most measurements of cold atomic clouds reduce to some record of the cloud's brightness, either under normal trapping conditions or when illuminated by a weak probe beam. In either case, the interpretation of those data is entirely dependent on the excited-state fraction of the trapped atomic cloud \cite{Steck:2010}.

Additionally, \hl{measurements} of cold-atom ionization cross-sections \cite{Wipple:2001,Prentiss:1988,Dinneen:1992} for cold quantum chemistry \cite{Hall:2011,Grier:2009,Rellergert:2011,Ratschbacher:2012, Sullivan:2012,Smith:2014,Goodman:2015} require accurate knowledge of the excited-state and ground-state populations. Experiments \cite{Sullivan:2012} and \textit{ab initio} calculations \cite{Cote:2000,McLaughlin:2014} show that knowledge of the electronic state of the reactants is necessary to determine these multi-channel reaction rates and corresponding branching ratios.

Surprisingly, $f_e$ is almost never directly measured, but is instead indirectly determined using an idealized two-level model.  \hl{For example, a model-dependent measurement was performed for a Rb MOT by Dinneen {\it et al.} \cite{Dinneen:1992}. More recently, Glover {\it et al.} \cite{Glover:2013} performed a model-dependent measurement on a Ne MOT.}

\hl{ The commonly used idealized two-level model is} based on the steady-state solution to the optical Bloch equations
\begin{equation}
f_e = \left(\frac{1}{2}\right)\frac{I/I_s}{1+I/I_s+(2\delta/\Gamma)^2},
\label{eq:fe}
\end{equation}
where $I$ is the total MOT laser intensity summed over the six beams, $\delta$ is the detuning from atomic resonance, and $\Gamma$ is the transition's natural linewidth \cite{Foot:2005,Steck:2010}. Here, the saturation intensity $I_s$ is consistent with the definition from Refs.~\cite{Foot:2005,Steck:2010}, e.g., for circularly polarized light, the theoretical saturation intensity is given by 
\begin{equation}
I_s = I_{s,\sigma} = \frac{\hbar \omega^3 \Gamma}{12\pi c^2},
\label{eq:Is}
\end{equation}
where $\omega$ is the angular frequency of the atomic transition, $\hbar$ is Planck's constant divided by $2\pi$, and $c$ is the speed of light in vacuum. By defining a saturation parameter
\begin{equation}
s \equiv \frac{I/I_s}{1+(2\delta/\Gamma)^2},
\label{eq:satp}
\end{equation}
we can write an alternative expression for Eq.~(\ref{eq:fe}) as
\begin{equation}
f_e = \frac{1}{2}\left (\frac{s}{1+s} \right).
\label{eq:fes}
\end{equation}

Despite the fact that MOTs have been commonly used in many cold atomic physics laboratories since the late 1980s \cite{Raab:1987}, only recently has there been any direct model-independent measurements of a MOT's excited-state population \cite{Shah:2007,Veshapidze:2015}. Moreover, these studies were limited in scope to two commonly used isotopes of rubidium (\ce{Rb}). Unexpectedly, those studies found that the simple two-level model has better predicting power than more sophisticated models \cite{Townsend:1995,Javanainen:1993}, if an experimentally determined effective saturation intensity $I_\mathrm{se}$ is used. These model-independent Rb measurements were able to precisely quantify this effective saturation intensity, demonstrating that its value remained constant over a wide range of trap settings.

References \cite{Shah:2007,Veshapidze:2015} found that the experimentally determined effective saturation intensity for \ce{^{87} Rb} is about 2.8 times larger than the circularly-polarized theoretical saturation intensity and 1.3 times larger than the isotropically-polarized theoretical saturation intensity \cite{Steck:2001}. In a sodium (Na) MOT, we expect the excited-state fraction of \ce{Na} to show an even greater departure from the idealized two-level system than was measured in \ce{Rb}. This is because the excited-state hyperfine structure in \ce{Na} is narrower, making repumper conditions more sensitive in \ce{Na} than in \ce{Rb}. 

\hl{For example, a rough calculation} shows that if we assume the cycling and re-pumper transition strengths are comparable and that $I \ll I_s$, the photon scattering rates\footnote{Here we assume $I\ll I_s$ for \hl{simplicity, but this approximation does not apply during the experiment. However, the qualitative conclusion that the Na has more sensitive repumper conditions than Rb remains the same, even if $I>I_\textrm{sat}$.}} \hl{ for} the type-I \ce{Na} or \ce{^{87} Rb} MOT's cycling transition ($F = 2$ to $F^\prime = 3$) and ``leakage'' transition ($F = 2$ to $F^\prime = 2$) are
\begin{align}
R_\mathrm{cycle} & \approx  \hl{\frac{\Gamma}{2} \frac{I/I_s}{1+(2\delta/\Gamma)^2}} \\
\mathrm{and}~~R_\mathrm{leak} &\approx \hl{\frac{\Gamma}{2} \frac{I/I_s}{1+[2(\Delta-\delta)/\Gamma]^2}},
\label{eq:Rscatt2} 
\end{align} 
respectively. Here, $\Delta$ is the splitting between the excited cycling $F^\prime=3$ state and the leakage $F^\prime=2$ state for the type-I MOT. By taking the ratio of these two rates and using the values from \mbox{Refs.~\cite{Steck:2001,Steck:2010}} for $\Gamma$ and $\Delta$, as well as assuming that $\delta \approx \Gamma/2$, we get an estimate that on average, the leakage excited-state is populated about once every \hl{60} cooling cycles for \ce{Na}, but only once every \hl{3900} cooling cycles for \ce{^{87} Rb}. Last, we would expect the effective saturation intensity for \ce{Na} or \ce{Rb} to be greater than the two-level model would predict since leakage to other states necessitates greater intensity to saturate the cycling transition.

In this paper, we demonstrate a new technique for performing a model-independent measurement of a type-I and type-II \ce{Na} MOT \cite{Tanaka:2007,Raab:1987} using a hybrid atom-ion trap apparatus \cite{Smith:2003,Smith:2005,Goodman:2012,Sivarajah:2012,Smith:2014,Goodman:2015,Wells:2017}. We will define the \ce{Na} atom's ``excited-state'' to be any hyperfine state in the $3 ^{2}P_{3/2}$ level of the D2 line. We compare our experimental results with a simple two-level model and find a clear departure. We extract a value for the effective saturation intensity for both the type-I and type-II MOTs.

Our paper is organized as follows: In Sec.~II, we briefly describe the salient points of our experimental apparatus. In Sec.~III, we discuss the method behind our model-independent measurement of $f_e$. In Sec.~IV and V, we discuss the results of this measurement, which includes a discussion of where and how the two-level model fails at a critical saturation intensity. In Sec.~VI, we conclude.
\begin{figure}
\centering
	\includegraphics[width=\linewidth]{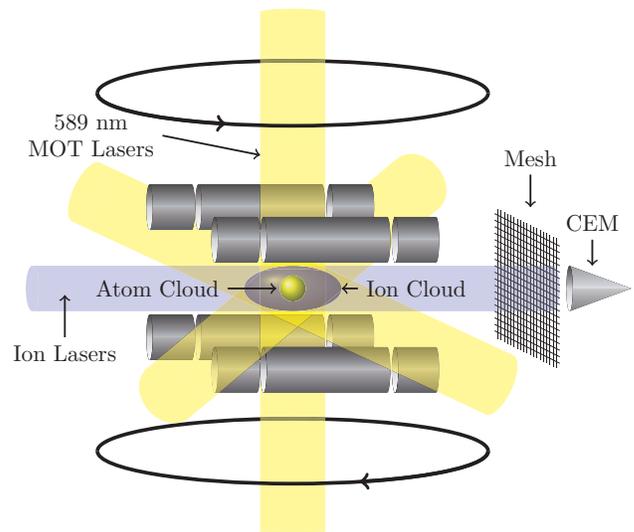}
	\caption{\label{fig:hybrid} (Color online) The hybrid trap consists of a concentric MOT and LPT. The MOT is made from 3 perpendicular retroreflected circularly polarized 589 nm beams and opposing magnetic-field coils shown at the top and bottom.  ``Ion lasers'' include a 405 nm laser to ionize excited sodium, and lasers for the creation and cooling of  \ce{Ca+} ions: 423 nm to excite ground-state calcium, 375 nm to ionize excited calcium, 397 nm to cool \ce{Ca+}, and 866 nm to repump \ce{Ca+}.  The lasers are not all used simultaneously. For the work presented here, we either work with \ce{Ca+} alone, for the purpose of calibrating the CEM, or we work with Na and \ce{Na+}. We apply an r.f. voltage to \hl{each diagonal pair of} central rods, \hl{where one diagonal pair is completely out of phase with the other.} A DC voltage \hl{is applied to} the eight end \hl{rods}. A typical ion cloud is depicted, concentric with the trapped atoms in the MOT. \hl{A destructive measurement of ion number is performed when we modulate the voltage of four end rods, and extract the ions through the mesh and into the CEM.}
	}
\end{figure}

\section{Apparatus}

A description of our apparatus can be found in previous works \cite{Goodman:2015,Wells:2017}. We will briefly describe our apparatus here, and the additional elements unique to this experiment.

Our hybrid trap consists of a concentric Na MOT and linear Paul trap (LPT), as seen in Fig.~\ref{fig:hybrid}. The MOT is vapor-loaded and made with six (three retro-reflected) beams of \hl{circularly-polarized} light tuned near the sodium D2 line. The radiation force in conjunction with magnetic-field gradients of $\approx 30\;\textrm{Gauss}/\textrm{cm}$ spatially confines and cools the atoms in the center of the trap. Sodium has two different hyperfine cycling transitions that can be used for trapping, resulting in two different types of MOTs, the type-I and type-II MOTs. \hl{A type-I MOT uses a cooling transition where $F'=F+1$; in sodium $F'=3$ and $F=2$. A type-II MOT uses a cooling transition where $F\geq F'$; in sodium $F' = 0$ and $F = 1$ \cite{Jarvis:2018,Oien:1997}.} Our type-I MOT typically holds $\sim 10^6$ atoms in steady-state at a temperature of $\approx 300\;\mu\textrm{K}$ and a peak density of $\sim 10^{10}\;\textrm{cm}^{-3}$. Our type-II MOT holds $\sim 10^7$ atoms in steady-state at a temperature of $\approx 2\;\textrm{mK}$ and a peak density of $\sim 10^{9}\;\textrm{cm}^{-3}$. 

\hl{We control the detuning of the cooling-laser by passing it through two acousto-optical modulators (AOMs) and locking the shifted laser to the peak of a known hyperfine transition in Na using saturation spectroscopy on a heated Na cell. From this, we use the modulation frequencies of the two AOMs to determine the detuning of the cooling-laser from the cycling transition resonance.} We use an electro-optic modulator (EOM) to add sidebands to our \hl{cooling} laser light. The EOM is driven with a frequency close to the ground-state spacing of sodium. One sideband is used to repump the Na atoms out of the dark ground-state, $F=1$ for the type-I MOT and $F=2$ for the type-II MOT. The EOM creates adjustable-strength sidebands up to 25\% of the intensity of the carrier.

\begin{figure}
	\includegraphics[width=\linewidth]{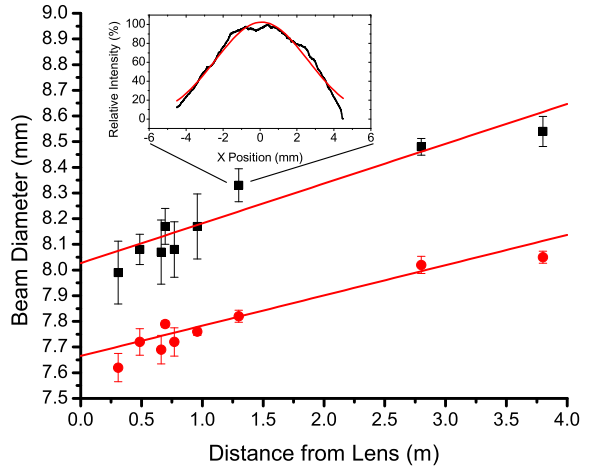}
	\caption{\label{fig:beamprofile} (Color online) The $1/e^2$ beam width is measured in the two transverse dimensions (red circles and black squares) along a range of distances from the EOM. The data were fit to Eq.~(\ref{eq:beamdiv}). This can be used to predict the size of each of the cooling-laser beams at the MOT, taking into account extra distance traveled by the retroreflection by each beam. The MOT is $\approx 1.3 \;\text{m}$ away from the lens, though a precise value is measured for each path independently. The inset shows the beam profile of a single measurement, and its corresponding fit to a Gaussian.}
\end{figure}

The EOM introduces some divergence to the laser beam, so in order to properly quantify the cooling-laser intensity at the MOT location, we measure the beam profile at several distances from the EOM using a ThorLabs BP209-VIS beam profiler. The beam profile approximates and is fit to a Gaussian $\textrm{TEM}_{00}$ spatial mode. With these data, we calculate the divergence of the beam by performing a two-parameter fit to the expected Gaussian $1/e^2$ beam width's dependence $w$ on the position along the beam $z$, given by

\begin{equation}
w(z)=w_0\sqrt{1+\left(\frac{z-z_0}{z_R}\right)^2},
\label{eq:beamdiv}
\end{equation} 
where $w_0$ is the beam waist, $z_0$ is the position of the beam waist, and $z_R\equiv\pi w_0^2/\lambda$ is the Rayleigh range. A fit to this equation is shown in Fig.~\ref{fig:beamprofile}, allowing us to extrapolate the size of each beam at the center of the MOT. We measure this daily to account for any day-to-day fluctuations in the uncertainty of the measurement of the beam size.

The segmented-electrode LPT makes up the second half of our hybrid apparatus, allowing us to spatially co-trap ions with cold MOT atoms. In this experiment, the LPT traps the ions created from photoionizing the MOT atoms. The photoionization (PI) is performed with a 405 nm diode laser for Na. The size of the 405 nm beam was determined using the same procedure as that of the 589 nm MOT beams. \hl{We can approximate the PI beam as having a uniform intensity distribution over the volume of the MOT, because the $1/e$ beam radius is at least twice that of the largest MOT we can create.} To trap the ions, we apply a $780\;\textrm{kHz}$ signal of $120\;\textrm{V}$ \hl{peak to peak} amplitude \hl{(relative to ground)} to \hl{each diagonal pair of central rods} in the LPT. \hl{More details of the LPT apparatus can be found in Ref.~\cite{Sivarajah:2013}}. \hl{The r.f.} creates a trap with a depth $\sim 1\;\textrm{eV}$, \hl{which} exceeds the MOT trap depth by two orders of magnitude. Additionally, the LPT's ion trapping volume has been determined through simulations and experiments \cite{Goodman:2015} to be about twice as large as the larger type-II MOT. Therefore, we can assume that all ions created from the MOT are initially trapped by the LPT. Since sodium ions have a \hl{Ne-like} closed electronic structure, we cannot use fluorescence detection methods to measure the ion trap population. Instead, we use a destructive measurement via a \textsc{megaspiraltron} Channeltron electron multplier (CEM) and a preamplifier, whose peak voltage output is proportional to the number of ions in the trap. The end electrodes of the LPT are gated from a trapping voltage configuration into a dipole configuration that \hl{rapidly} extracts the ions from the trapping region and into the CEM, which is coaxial with the LPT.

In order to perform a calibration of our CEM, we \hl{use laser cooling to} create a Coulomb crystal of \ce{Ca+} which we extract and detect with the CEM. Since the \ce{Ca+} crystal \hl{fluoresces} with a 397 nm laser \cite{Toyoda:2001}, we can image the crystal with a CCD camera before extraction and directly count the number of ions, giving us an absolute calibration on our CEM \cite{Grier:2009}. We calibrate the CEM with linear crystals of between one and twelve ions at a CEM cone-voltage setting of \mbox{2250 V}. In a separate measurement, we determine the ratio between the sensitivity used for the calibration, 2250 V, and the sensitivities used for the expeirment. The CEM gain is exponentially dependent on the cone-voltage setting \cite{Goodman:2015}, as is expected when the CEM is not saturated. Because of differing total PI rates, we perform the experiment at $1750\;\textrm{V}$ (higher sensitivity) and $1500\;\textrm{V}$ (lower sensitivity) for the type-I and II MOTs, respectively. We find the ratio between different CEM sensitivities by repeatedly loading a similar number of \ce{Ca+} ions into the LPT and extracting at each CEM setting independently. If we know the absolute calibration at one setting and the ratio between settings, then we can determine the absolute calibration at any setting. \hl{Our CEM calibration factor was determined to be $\kappa_\mathrm{CEM}=1560\pm 110\;\textrm{ions/V}$ for the 1750 V setting, and $\kappa_\mathrm{CEM}=54800\pm3700\;\textrm{ions/V}$ for the 1500 V setting.} With our calibrated CEM, our destructive measurement of sodium ions yields a direct measurement of the number of atoms photoionized within a given amount of time.

\section{Experiment}

A model-independent measurement of $f_e$ can be made by comparing two methods of measuring the number of ions created from the MOT via PI within our hybrid trap: directly, with our LPT and calibrated CEM, and indirectly by monitoring the change in MOT fluorescence when exposed to the PI laser. We will begin with a discussion of the latter method. The total PI rate of the MOT, $\gamma_\mathrm{pi}$, which is proportional to the MOT's excited-state fraction, is defined as
\begin{equation}
\gamma_\mathrm{pi}=\frac{\sigma_\mathrm{pi} f_e I_\mathrm{pi}}{h \nu_\mathrm{pi}} \equiv \zeta I_\mathrm{pi},
\label{eq:gpi}
\end{equation}
where $\sigma_\mathrm{pi}$ is the PI cross section, $I_\mathrm{pi}$ is the intensity of the PI laser, and $h \nu_\mathrm{pi}$ is the energy per PI photon \cite{Dinneen:1992,Wipple:2001,Petrov:2000,Preses:1985}. 

We operate our MOT in the temperature-limited regime \cite{Townsend:1995, Wipple:2001,Goodman:2015}, where the volume of the MOT $V_\mathrm{MOT}$ remains constant during loading, and thus the temperature remains constant since the two are proportional. Meanwhile, the MOT density $n_\mathrm{MOT}$ increases linearly with increasing atom population $N_a$. Collisions between two MOT atoms lead to a quadratic two-body loss rate $\beta n_\mathrm{MOT}$ \cite{Prentiss:1988}. Collisions with constant density uncooled background \ce{Na} atoms result in a linear loss rate $\gamma_b$.  We model the MOT loading behavior with a non-linear rate equation
\begin{equation}
\frac{dN_a}{dt} = L_\mathrm{MOT} - \gamma_t N_a - \frac{\beta}{V_\mathrm{MOT}} N_a^2,
\label{eq:dtMOT}
\end{equation}
where $ L_\mathrm{MOT}$ is the constant rate at which atoms are loaded into the MOT, and $\gamma_t$ is the total single-body linear loss rate \cite{Wipple:2001}. If the only single-body loss rate is due to background gas collisions, then $\gamma_t = \gamma_b$. The general solution to Eq.~(\ref{eq:dtMOT}) is
\begin{equation}
N_a(t)=\frac{2L_\mathrm{MOT} \left ( 1-e^{-\gamma_e t} \right )}{\gamma_e + \gamma_t + \left ( \gamma_e - \gamma_t \right ) e^{-\gamma_e t}},
\label{eq:MOTsol}
\end{equation}
where 
\begin{equation}
\gamma_e = \sqrt{\gamma_t^2+\frac{4 \beta L_\textrm{MOT}}{V_\textrm{MOT}}}.
\label{eq:ge}
\end{equation}

The steady-state atom population $\tilde{N}_a$ can be found by taking the limit of Eq.~(\ref{eq:MOTsol}) as $t\rightarrow \infty$, which yields 
\begin{equation}
\tilde{N}_a = \frac{2L_\mathrm{MOT}}{\gamma_t + \sqrt{\gamma_t^2 + \frac{4 \beta L_\mathrm{MOT}}{ V_\mathrm{MOT}}}}.
\label{eq:Nass}
\end{equation}
\hl{To convert from atom units to PMT signal (voltage) units we use the} energy per 589 nm photon $E_\mathrm{MOT}$, the known detector collection efficiency factor related to the fraction of the total solid angle imaged on to the PMT $\eta$, and most importantly, the excited-state fraction of atoms $f_e$. \hl{The geometric collection efficiency $\eta = (1.59\pm 0.05)\times 10^{-3} $ remains constant throughout the experiment. The absolute calibration of the PMT at the MOT wavelength is captured in the variable $c_\mathrm{PMT}$. This calibration was determined by shining a weak laser directly into the PMT, giving us the ratio of signal voltage to incident 589 nm laser power. We can express $L_\textrm{MOT}$ in terms of the PMT measured loading rate $L_\textrm{PMT}$ (volts per second) as}
\begin{equation}
L_\mathrm{MOT}=\left ( \frac{1}{\hl{\eta c_\mathrm{PMT}} E_\mathrm{MOT} \Gamma} \right ) \frac{L_\mathrm{PMT}}{f_e} \equiv \frac{\kappa_\mathrm{PMT}}{f_e}L_\mathrm{PMT},
\label{eq:LMOT}
\end{equation}
where we have combined the prefactors into an overall PMT calibration, $\kappa_\textrm{PMT}$ \hl{multiplied by the PMT measured loading rate $L_\textrm{PMT}$. In a typical experiment for the type-I MOT, $\kappa_\textrm{PMT}=(7.19\pm 0.23)\times10^5\;\textrm{atoms/V}$. Clearly, it is also true that}
\begin{equation}
\hl{N_\mathrm{MOT}=\frac{\kappa_\mathrm{PMT}}{f_e}N_\mathrm{PMT},}
\end{equation}
\hl{where $N_\mathrm{PMT}$ is the PMT voltage signal proportional to the excited atom number.} For convenience, we group several of the constants into a directly measured MOT \hl{loss rate $D$}, which is equivalent to $\beta\kappa_\mathrm{PMT}/(V_\mathrm{MOT}f_e)$. Rewriting Eq.~(\ref{eq:MOTsol}) in terms of parameters we experimentally measure yields

\begin{equation}
\hl{N_\mathrm{PMT}(t)=\frac{2L_\mathrm{PMT} \left ( 1-e^{-\gamma_e t} \right )}{\gamma_e + \gamma_t + \left ( \gamma_e - \gamma_t \right ) e^{-\gamma_e t}}},
\label{eq:MOTsolfit}
\end{equation}
where we have rewritten $\gamma_e$ as 
\begin{equation}
\gamma_e = \sqrt{\gamma_t^2+4 D L_\textrm{PMT}},
\label{eq:ge2}
\end{equation}

\begin{figure}
	\includegraphics[width=\linewidth]{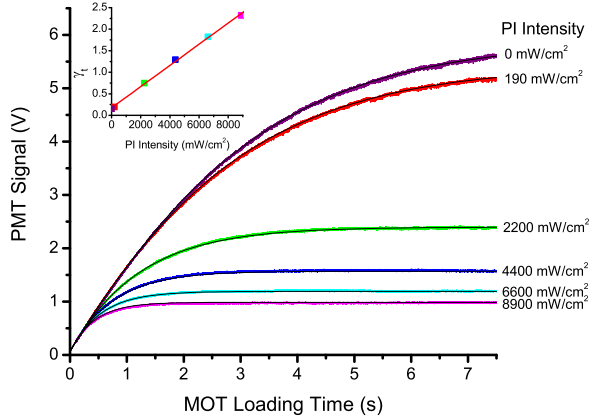}
	\caption{\label{fig:motload} (Color online) The MOT \hl{loss rate} changes as a function of PI intensity $I_{pi}$. Each curve is fit to Eq.~(\ref{eq:MOTsolfit}). The inset shows the fitted value of the total \hl{loss rate} $\gamma_t$ vs. the PI intensity $I_\mathrm{pi}$. The the slope of the linear fit within the inset is equivalent to $\zeta$ and the $y$ intercept is equivalent to $\gamma_b$, from Eq.~(\ref{eq:gammat}).}
\end{figure}

When the MOT is also experiencing PI, there is an additional one-body loss rate $\gamma_\mathrm{pi}$, which increases the total loss rate
\begin{equation}
\gamma_t = \gamma_b + \gamma_\mathrm{pi}= \gamma_b+\zeta I_\mathrm{pi}.
\label{eq:gammat}\end{equation}
By fitting the MOT loading curves to Eq.~(\ref{eq:MOTsolfit}), \hl{we obtain fit values for $L_\textrm{PMT}$, $D$, and $\gamma_t$. The fit values of $L_\textrm{PMT}$ and $D$ do not change with $I_\textrm{pi}$, but Eq.~(\ref{eq:gammat}) suggests that $\gamma_t$ changes linearly with $I_\textrm{pi}$.} A fitted slope and $y$-intercept of a $\gamma_t$ vs.~$I_\mathrm{pi}$ scatter plot will yield $\zeta$ and $\gamma_b$, respectively. Figure \ref{fig:motload} shows a plot of the PMT MOT-loading data with a corresponding fit to Eq.~(\ref{eq:MOTsolfit}). A representative plot of $\gamma_t$ vs. $I_{pi}$ can be seen in the inset, fit to Eq.~(\ref{eq:gammat}).

\begin{figure}
	\includegraphics[width=\linewidth]{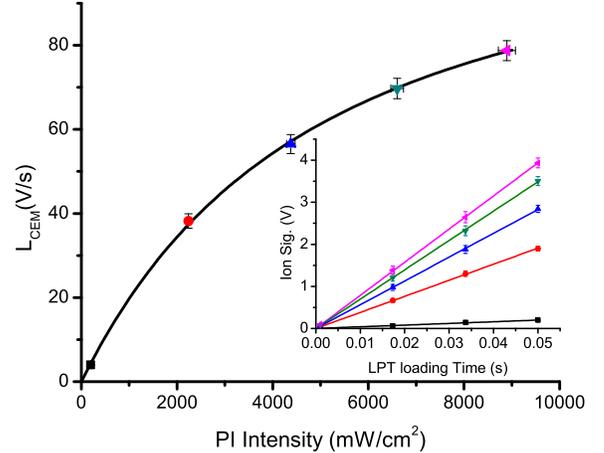}
	\caption{\label{fig:fefit} (Color online) A typical curve of $L_\textrm{CEM}$ vs. $I_\textrm{pi}$ is shown. A one-parameter fit with Eq.~(\ref{eq:indfe}) gives a model-independent value for $f_e$. All other variables in Eq. (\ref{eq:indfe}) are directly measured independently. Each value for $L_\mathrm{CEM}$ is calculated from a linear fit of a CEM loading curve, shown in the inset. Each loading curve corresponds to a specific PI intensity, which is plotted on the corresponding $L_\textrm{CEM}$ vs. $I_\textrm{pi}$ curve with the same data point shape and color. The PI intensity was measured before each LPT loading curve. The uncertainty in PI intensity comes from the standard deviation of this set of intensities.}
\end{figure}

By suddenly turning on the LPT while the MOT is in steady-state and subjected to PI radiation, we can load the LPT for a variable duration $t_\mathrm{load}$. We use the \hl{calibrated} CEM to measure the number of \ce{Na+} ions created during this loading time. As discussed earlier, we assume that every ion created from the MOT becomes an ion loaded into the LPT. The loading rate becomes

\begin{equation}
L_I =\tilde{N}_a \gamma_\mathrm{pi} = \tilde{N}_a(I_\mathrm{pi})\,\zeta(I_\mathrm{MOT})\,I_\mathrm{pi},
\label{eq:NI}
\end{equation}
where we have emphasized that $\tilde{N}_a$ is a function of $I_\mathrm{pi}$, due to its dependence on $\gamma_t$ in Eq.~(\ref{eq:Nass}) and that $\zeta$ is a function of the total cooling-laser intensity $I_\mathrm{MOT}$, due to its dependence on $f_e$ in Eq~(\ref{eq:gpi}). We have verified experimentally that nearly all of the ions loaded into the LPT come from the MOT and not the excited uncooled background Na vapor. This is a consequence of the MOT being several orders of magnitude more dense than the background gas. For small values of $t_\mathrm{load}$, as compared to the time it takes the LPT to saturate, we expect $N_I = L_I t_\mathrm{load}$, making $L_I$ extractable from plots of $N_I~vs.~t_\mathrm{load}$, as previously shown in Refs.~\cite{Goodman:2015,Wells:2017} and shown here in the inset of Fig.~\ref{fig:fefit}. If the CEM is calibrated, then $L_I$ can be expressed in units of ions per second, i.e.,
\begin{equation} 
L_I = L_\mathrm{CEM} \kappa_\mathrm{CEM},
\label{eq:LIcal}
\end{equation}
where $L_\mathrm{CEM}$ is the loading rate measured in CEM signal-voltage per second and $\kappa_\mathrm{CEM}$ is the calibration for the number of ions trapped per CEM signal volt.

\begin{figure*}
	\includegraphics[width=.5\linewidth]{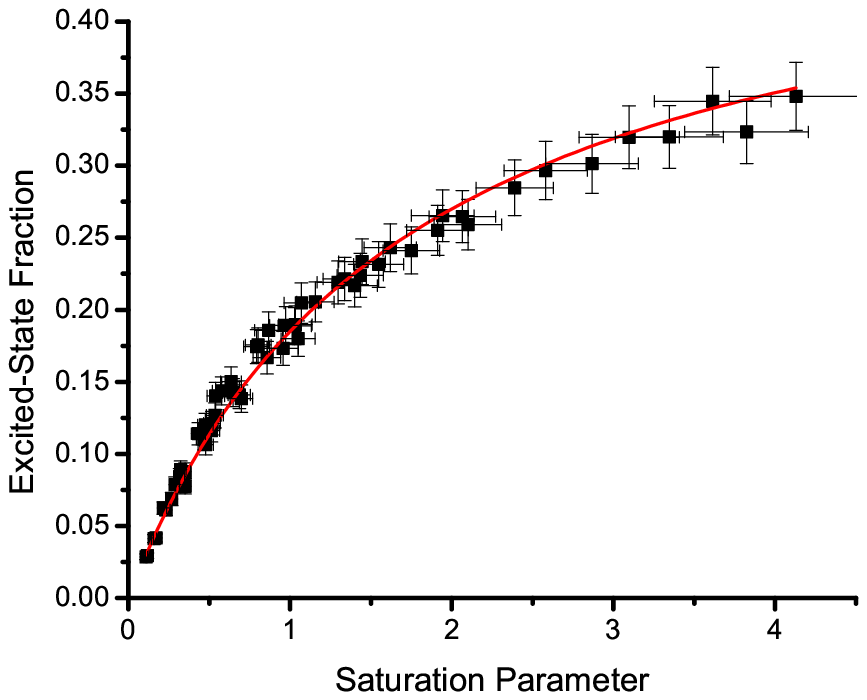}\includegraphics[width=.5\linewidth]{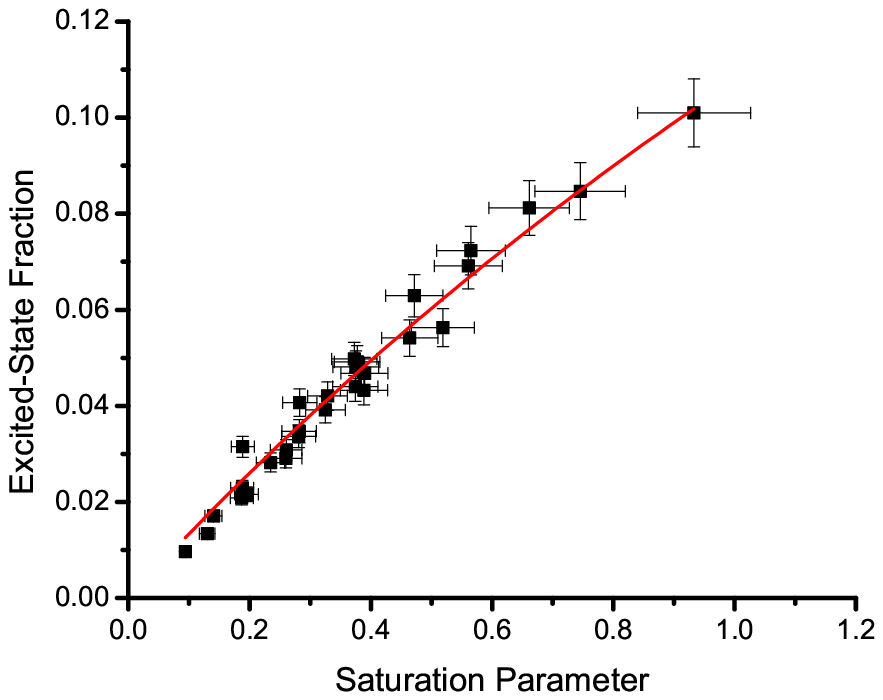}
\caption{\label{fig:alldata} (Color Online) The model-independent measurement of excited-state fraction $f_e$ is shown vs. saturation parameter $s$ for the type-I (left) and type-II (right) MOTs. The solid (red) line shows a fit to Eq.~(\ref{eq:fes2}) with only one free-fitting parameter, the ratio $I_\textrm{s}/I_\textrm{se}$. The data shown in both the type-I and type-II plots were taken over a wide range of MOT settings to demonstrate that the fitted curve is universal. To begin with, we considered a range of magnetic-field gradients from $25\;\textrm{Gauss}/\textrm{cm}$ to $40\;\textrm{Gauss}/\textrm{cm}$. The repump intensity in each case was varied between $5\%$ and $25\%$ of the total cooling-laser intensity. Finally, the data shown for the type-I MOT were taken for 5 different cooling-laser detunings between $7\;\textrm{MHz}$ and $18\;\textrm{MHz}$, and for \hl{4} different cooling-laser detunings between \hl{$10\;\textrm{MHz}$} and $22\;\textrm{MHz}$ for the type-II MOT.
}
\end{figure*}
Substituting Eq.~(\ref{eq:LIcal}) into the left-hand-side of Eq.~(\ref{eq:NI}) and substituting Eq.~(\ref{eq:Nass}) into the right-hand-side of Eq.~(\ref{eq:NI}) gives
\begin{widetext}
\begin{equation}
L_\mathrm{CEM} = \frac{1}{f_e} \left (\frac{\kappa_\mathrm{PMT}}{\kappa_\mathrm{CEM}} \right ) \frac{2L_\mathrm{PMT} \zeta I_\mathrm{pi}}{\gamma_b+\zeta I_\mathrm{pi} + \sqrt{(\gamma_b+\zeta I_\mathrm{pi})^2 + 4 D L_\mathrm{PMT}}},
\label{eq:indfe}
\end{equation}
\end{widetext}
where we have also substituted Eq.~(\ref{eq:LMOT}) for $L_\mathrm{MOT}$ in Eq~(\ref{eq:Nass}). Except for $f_e$, all of the parameters in Eq.~(\ref{eq:indfe}) are directly determined experimentally: $L_\textrm{PMT}$ and $D$ are determined by a fit to Eq.~(\ref{eq:MOTsolfit}), $\gamma_b$ and $\zeta$ were determined by a fit to Eq.~(\ref{eq:gammat}), and $\kappa_\textrm{PMT}$ and $\kappa_\textrm{CEM}$ were determined directly and remain constant throughout the experiment. Therefore, a plot of $L_\mathrm{CEM}~vs.~I_\mathrm{pi}$ has a single fitting parameter, which is the model-independent $f_e$ at a fixed $I_\mathrm{MOT}$. A typical data set for the type-I MOT is shown in Fig.~\ref{fig:fefit}.

\hl{To determine the uncertainty in $f_e$, we average the propagated uncertainty from each data point in Fig. \ref{fig:fefit}.} Some of the variables in the fit are correlated. However, an analysis reveals that these errors were much smaller than the error in $\kappa_\mathrm{CEM}$ \hl{ and $\kappa_\mathrm{PMT}$}, which are by far the dominant sources of error in this measurement. Thus, the correlated error \hl{correction was} not included in the final analysis for each $f_e$ data point.

Last, by separately fitting a family of $L_\mathrm{CEM}~vs.~I_\mathrm{pi}$ plots for different values of $I_\mathrm{MOT}$, we can generate a model-independent plot of $f_e~vs.~s$, as seen in Fig.~\ref{fig:alldata}. In order to model this behavior with the effective two-level model, we must substitute \hl{an effective saturation intensity $I_\mathrm{se}$ for $I_\mathrm{s}$}. To do this, we rewrite Eq.~(\ref{eq:fes}) as
\begin{equation}
f_e = \frac{1}{2}\left (\frac{s I_\mathrm{s}/I_\mathrm{se}}{1+s I_\mathrm{s}/I_\mathrm{se}} \right),
\label{eq:fes2}
\end{equation}
where the ratio $(I_\mathrm{s}/I_\mathrm{se})$ is a free fitting-parameter. Each data point's $s$ value is calculated using the isotropically polarized theoretical value for $I_\mathrm{s}=13.4144(45)\;\text{mW}/\text{cm}^2$ \cite{Steck:2010}. Last, with the theoretical value for $I_\mathrm{s}$ and fitting result for the ratio  $I_\mathrm{s}/I_\mathrm{se}$, we solve for $I_\mathrm{se}$.

\section{Two-Level Fit}
For low cooling-laser intensity, we see that our data follow the simple predictive two-level model, regardless of the chosen value for the cooling-laser detuning, repump intensity, or magnetic-field gradient, as shown in Fig.~\ref{fig:alldata}. Here, we only consider repump intensities which sufficiently saturate the repump transition, leaving effectively no population in the dark ground state. For the type-I MOT, the fit in Fig.~\ref{fig:alldata} predicts an effective saturation intensity \hl{$I_\mathrm{se}=22.9\pm5.1\;\text{mW}/\text{cm}^2$. To determine the uncertainty in $I_\textrm{se}$ we calculate the propagated uncertainty predicted by each data point and then average those uncertainties, which show little variance. The average propagated uncertainty is much larger than the purely statistical uncertainty of 1.3\%, determined using the standard deviation of the mean of $I_\textrm{se}$ nominal values.} Our experimental result is approximately 1.7 times larger than the theoretical isotropically-polarized saturation-intensity reported in Ref. \cite{Steck:2010}. We find that there is some critical intensity, above which $f_e$ becomes systematically dependent on detuning, repump intensity, or MOT magnetic-field gradient. The plots in Fig.~\ref{fig:alldata} only include data up to this point. The critical intensity for our Na MOT is dependent on the cooling-laser detuning, as one might expect. For the type-I MOT, the lowest measured critical total intensity of the six MOT beams was about $100\;\text{mW}/\text{cm}^2$ which corresponded with the greatest detuning that was tested. Thus, the effective two-level model can accurately predict the excited-state fraction for typical type-I MOT operating conditions. Determination of the critical intensity value is the subject of Sec. V.

\begin{figure}
	\includegraphics[width=\linewidth]{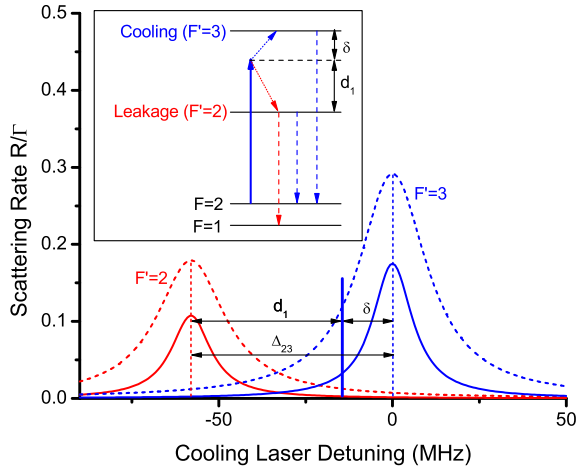}
	\caption{\label{fig:statemix} (Color Online) The scattering rate given in Eq.~(\ref{eq:scatterrate}) is plotted as a fraction of the decay rate $\Gamma$ against the cooling-laser detuning for the states discussed for the type-I MOT. The solid ($I/I_{sat}=1$) and dashed ($I/I_\textrm{sat}=5$) curves show the effects of power broadening on the scattering rate. The cooling-laser is shown as a solid blue vertical line, a frequency of $\delta$ detuned from the $F'=3$ state. The frequency difference between the $F'=2$ and $F'=3$ states is labeled as $\Delta_{23}$. The inset shows the level structure, as well as the pumping due to the cooling-laser into the cooling (blue dotted arrow) and leakage (red dotted arrow) states. As the cooling-laser intensity is increased, the probability of leakage is enhanced.
	}
\end{figure}

\begin{figure*}
	\includegraphics[width=.5\linewidth]{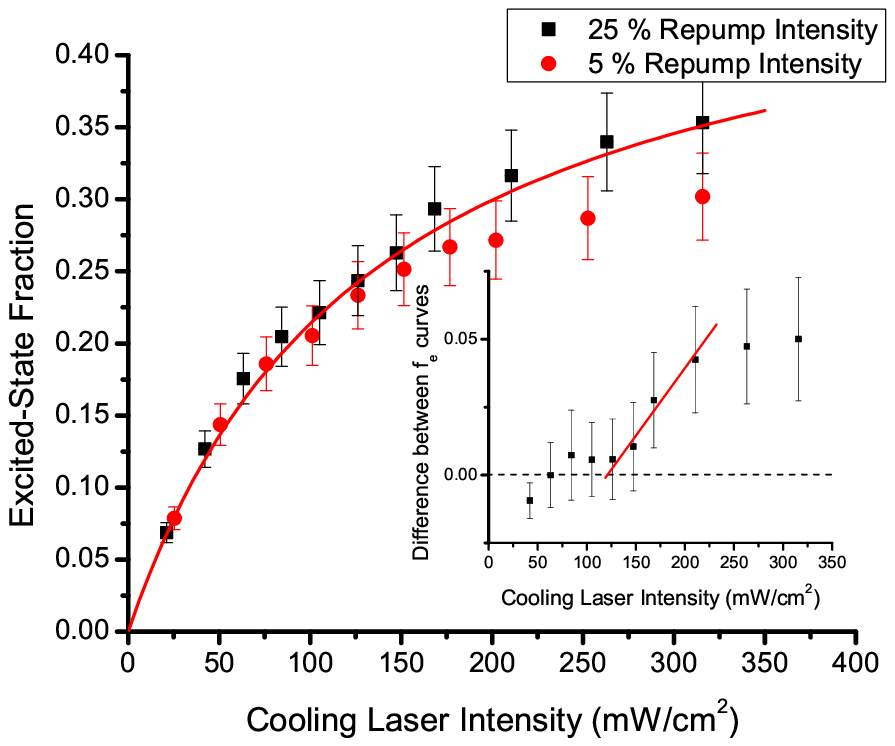}\includegraphics[width=.5\linewidth]{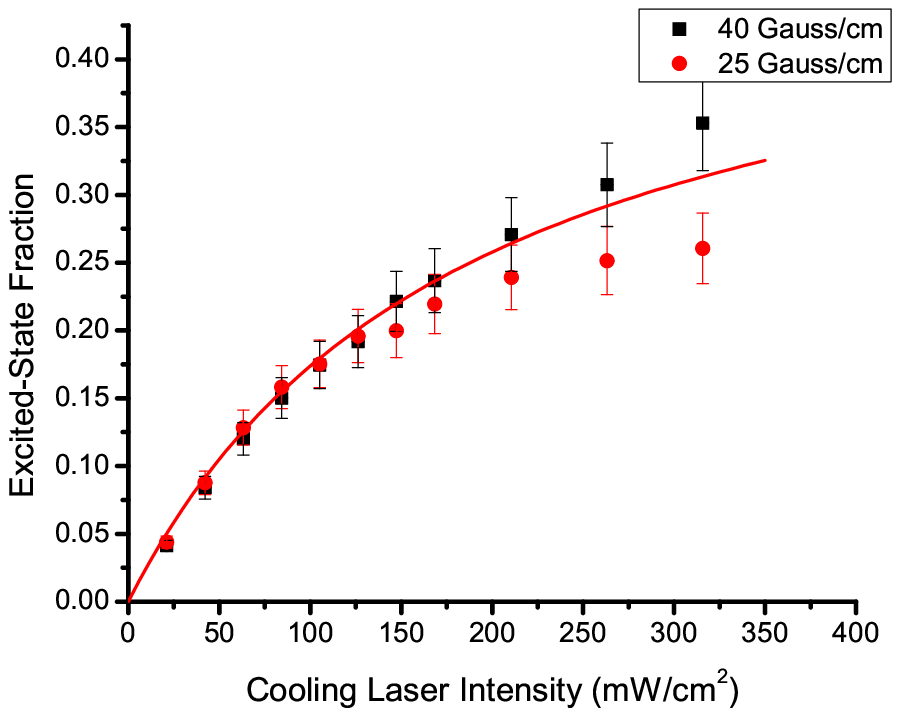}
\caption{\label{fig:repump} (Color Online) The representative behavior of $f_e$ in the type-I MOT is shown as we change the ratio of repump to cooling intensity (left) and magnetic-field gradient (right). Below some critical intensity (in both cases, approximately $150\;\textrm{mW}/\textrm{cm}^2$), these points seem to fall along the same universal curve. Above that critical intensity, $f_e$ becomes dependent on the repump-laser intensity ratio as well as the magnetic-field gradient. We estimate the point at which the two curves split by interpolating the points in one data set and comparing the difference in $f_e$ to the other data set. Once the difference is greater than the error in the measurement, we fit the deviation to extrapolate backwards and find $I_c$, shown in the inset of the left figure.  The fit included is the same as in Fig.~\ref{fig:alldata}, and is included to guide the eye.}
\end{figure*}

A similar analysis was done on the type-II MOT, as seen in the right side of Fig.~\ref{fig:alldata}. However, in the type-II MOT, there is a much smaller range of intensities for which $f_e$ is independent of detuning, repump intensity, and magnetic-field gradient. We were able to fit these data, yielding a saturation intensity of \hl{\mbox{$I_\mathrm{se}=49\pm11\;\textrm{mW}/\textrm{cm}^2$} with a separate statistical uncertainty of 1.4\%.} Unfortunately, there were certain detunings where $f_e$ was dependent on detuning, repump intensity, and magnetic-field gradient, regardless of intensity. \hl{Data for these small detunings ($\delta < 10\;\textrm{MHz}$) were not included in the fit for saturation intensity, shown in Fig.~ \ref{fig:alldata}.} Consequently, the type-II MOT's $f_e$ must be measured directly with a model-independent method, if an accurate excited-state fraction is desired for a type-II MOT.

\section{State-Mixing Behavior}

The region above the critical trapping-laser intensity where $f_e$ systematically depends on specific apparatus settings, in a manner that is not captured by the simple two-level model, is problematic for the greater experimental community. Therefore, we must analyze the mechanism behind the model breakdown and try to predict when the two-level model is no longer valid.

When the cooling-laser intensity is low, the leakage state's linewidth can be considered narrow enough that atoms primarily follow the cycling transition. However, as the intensity of the cooling-laser increases, power broadening of the leakage state by the cooling-laser results in more efficient population transfer out of the cycling transition, shown in Fig.~\ref{fig:statemix}.

We will qualitatively discuss this effect in the context of the type-I MOT. Once atoms are in the leakage $F'=2$ state, they can fall to the $F=1$ ground-state. Since the repump-laser couples the $F=1$ ground-state to the leakage state, our steady-state population in the leakage state becomes significant and dependent on the coupling of the cooling-laser to the leakage state $F=2\to F'=2$, the coupling of the repump-laser to the leakage state $F=1\to F'=2$, and the spontaneous decay out of the leakage state into both ground-states. In the case where the coupling into the leakage state is strong compared to the decay out of it, we see an enhancement in $f_e$ over the two-level model, since our measurement of $f_e$ includes both the $F'=2$ and 3 states. Alternatively, when atoms decay out of the leakage state more efficiently than they can be repumped, we see a decrease in $f_e$ below the two-level model. This results in a decrease in the overall excited-state population of the MOT. The magnetic-field gradient and repump-laser intensity both affect the coupling of the repump-laser into the excited leakage state, and as a result will change the steady-state populations in the total excited-state hyperfine manifold. In both cases, we only make quantifiable predictions in a limited regime of intensities, seen in Fig.~\ref{fig:repump}.

In their studies of the Rb MOTs, Shah and Veshapidze \cite{Shah:2007,Veshapidze:2015} found that the $f_e$ in their MOT followed the two-level model up to a saturation parameter of $s=1.25$ regardless of repump intensity, magnetic-field gradient, or detuning settings. In a Na MOT however, there is a critical intensity\footnote{Note that while saturation parameter and intensity are proportional, we observe an effect which depends on the detuning and intensity, so we discuss our deviation from the two-level model as a critical intensity rather than a critical saturation parameter.} $I_c$, above which $f_e$ diverges from the two-level model in a manner that depends on the particular repump intensity and/or magnetic-field gradient settings, as seen in Fig.~\ref{fig:repump}. Specifically, we see $f_e$ increase/decrease as a function of increased/decreased repump intensity and magnetic-field gradient for intensities $I>I_\textrm{c}$. Both of these behaviors are consistent with a state-mixing effect.
 
In order to model the onset of significant state-mixing for either MOT, we will introduce the power broadened photon absorption rate per ground-state atom involved in the cycling transition into the leakage hyperfine state (e.g., $F' = 2$ in the type-I MOT or the $F' = 1$ state in the type-II MOT) due to the cooling-laser as
\begin{equation}
R(d_n,I)=\left(\frac{\chi\Gamma}{2}\right)\frac{(I/I_\textrm{se})}{1+4(d_n/\Gamma)^2+(I/I_\textrm{se})},
\label{eq:scatterrate}
\end{equation}
where $d_n$ is the detuning of the cooling-laser to the leakage state for the type-$n$ MOT, and the hyperfine transition strength factor $\chi=1/4$ (type-I) and $\chi=5/12$ (type-II) \cite{Steck:2010}. This rate approximation doesn't account for stimulated emission, since we are working in the limit of low population in the leakage state. When working with the type-I MOT, the excited-state spacing between the cooling and leakage states is $\Delta_{23}$. If our cooling-laser detuning is $\delta$, then the difference in frequency to the leakage state is $d_1=\Delta_{23}+\delta$ as shown in Fig.~\ref{fig:statemix}. For the type-II MOT, the cooling transition is lower in frequency than the leakage transition, so $d_2=\delta-\Delta_{01}$. 

By comparing the scattering rate of the leakage state to the decay rate out of the leakage state, we can determine the excitation rate which causes significant population to be transferred into the leakage state, thus violating the two-level assumption. We assume that this happens when the rate becomes some critical fraction $f_c$ of the spontaneous decay rate out of the leakage state, $\Gamma$. For a fixed detuning, we can determine the critical cooling-laser intensity $I_c$, above which the two-level model no longer holds. At this critical intensity, we set $f_c \Gamma=R(d_n,I_c)$, which gives us
\begin{equation}
f_c\Gamma=\left(\frac{\chi\Gamma}{2}\right)\frac{(I_c/I_\textrm{se})}{1+4(d_n/\Gamma)^2+(I_c/I_\textrm{se})}.
\end{equation}
Solving this equation for the critical intensity, we see that
\begin{equation}
I_c(d_n,f_c)=\frac{2I_\textrm{se}(\Gamma^2+d_n^2)}{\Gamma^2}\frac{f_c}{2f_c-\chi}
\label{eq:fitIcrit}
\end{equation}

 \begin{figure*}
	\includegraphics[width=.5\linewidth]{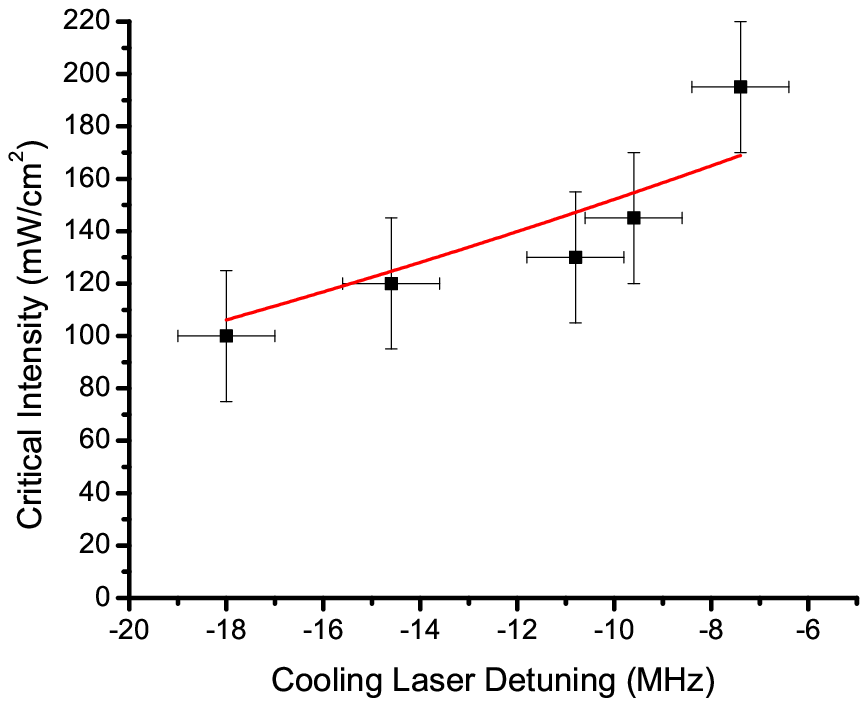}\includegraphics[width=.5\linewidth]{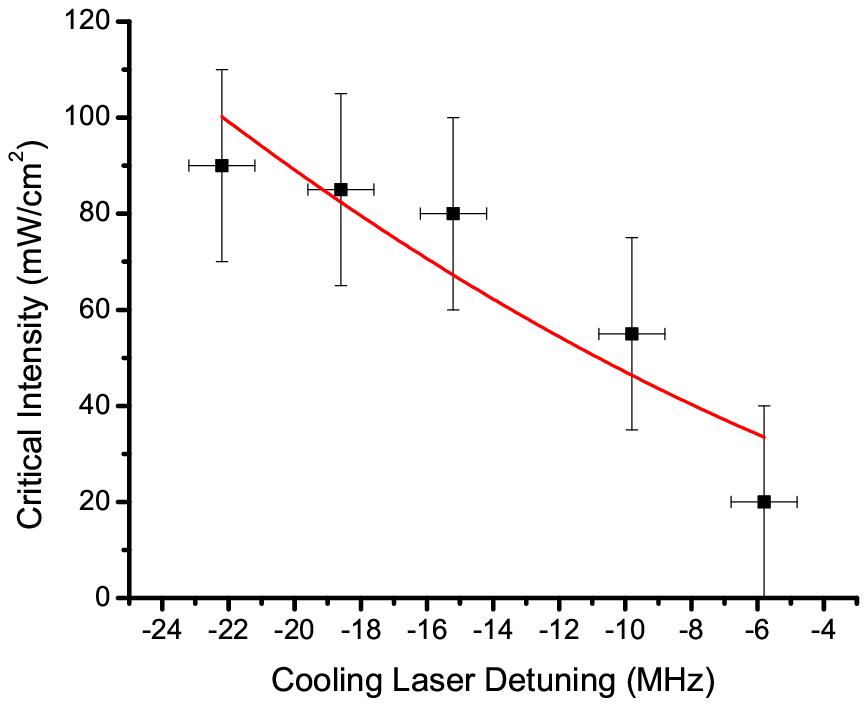}
\caption{\label{fig:critintens} The critical intensity \hl{(above which $f_e$ diverges as a function of cooling-laser intensity)} is shown as a function of the cycling transition's detuning from atomic resonance for the Na type-I MOT (left) and the type-II MOT (right) is shown. Each data point is calculated from a plot of $f_e$ vs. cooling-laser intensity. The error of each $I_c$ is determined through the fit of the differences in $f_e$. The data in the left and right plot fit with Eq.~(\ref{eq:fitIcrit}). These fits use the defined detunings $d_1(\delta)=\Delta_{23}+\delta$ and $d_2(\delta)=\delta-\Delta_{01}$ for the type-I and type-II MOTs, respectively.}

\end{figure*}

Using this function with $f_c$ as a single fitting parameter, we obtain the fits in Fig.~\ref{fig:critintens} and find $f_c=0.80(4)\%$ and $f_c=0.72(6)\%$ for the type-I and II MOTs, respectively. Since this state-mixing effect is only dependent on the rate into the leakage state, $f_c$ should be consistent across MOTs, since the hyperfine transition strength was accounted for. This is consistent with our findings. For comparison, to reach a fractional excitation rate of $0.72\%$ of the leakage state in a $^{87}\textrm{Rb}$ MOT, with a cooling-laser detuning of $d=\Gamma/2$ and saturation intensity of $9.2\;\text{mW}/\textrm{cm}^2$ \cite{Shah:2007}, would require a cooling-laser intensity $I_c\approx 4000\;\textrm{mW}/\textrm{cm}^2$. This is far outside the range of typical experimental parameters, explaining why previous studies did not observe a similar effect.

\section{Conclusions}

We have demonstrated a novel method to directly measure the excited-state fraction in a Na MOT using an ion-neutral hybrid trap. We found that for low cooling-laser intensities, the Na MOT follows a two-level model with an effective saturation intensity \hl{$I_\mathrm{se}=22.9\pm5.1\:\textrm{mW}/\textrm{cm}^2$ for the type-I Na MOT and $I_\mathrm{se}=49\pm11\;\textrm{mW}/\textrm{cm}^2$} for the type-II MOT. These two saturation intensities represent significant departures from the theoretically predicted saturation intensity reported in Ref.~\cite{Steck:2010} of $13.4144(45)\;\text{mW}/\text{cm}^2$.

At large enough intensities, we have observed a departure from the two-level model as a function of cooling-laser detuning, repump-laser intensity, and magnetic-field gradient. The critical cooling-laser intensity required to observe this departure changes as a function of cooling-laser detuning as expected. We find that the critical intensity for the type-I MOT is much higher than for the type-II MOT, \hl{due to the much smaller energy difference between the excited-state hyperfine levels corresponding to the cooling and leakage states}. This means that the two-level model is predictive over the typical operating parameters for the type-I MOT. We have implemented a \hl{model in Sec. V to predict when the leakage state is efficiently excited by the cooling-laser}. This model, along with the behavior of the excited-state fraction for high cooling-laser intensity suggests that the deviation from the predictive model is due to state-mixing between the cycling and leakage hyperfine states, caused by power broadening from the cooling-laser.

\section{Acknowledgements}

We would like to acknowledge support from the NSF under Grant No. PHY-1307874.

\end{document}